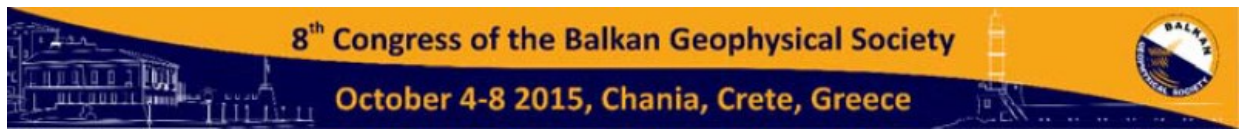

# Lithosphere-asthenosphere system in the Mediterranean region in the framework of polarized plate tectonics

*Reneta B. Raykova[1], Giuliano F. Panza[2,3,4,5], and Carlo Doglioni[6]*
[1] Sofia University "St. Kl. Ohridski", Faculty of Physics, Department of Meteorology and Geophysics, Blvd. J. Bauchier 5, Sofia; rraykova@phys.uni-sofia.bg
[2] University of Trieste, Department of Mathematics and Geosciences, via E. Weiss 4, Trieste, Italy; panza@units.it
[3] The Abdus Salam International Centre for Theoretical Physics, SAND Group, Trieste – Italy
[4] International Seismic Safety Organization (ISSO) - www.issoquake.org
[5] Institute of Geophysics, China Earthquake Administration, Beijing
[6] Department of Earth Sciences, University "Sapienza" of Rome, Italy, carlo.doglioni@uniroma1.it

## Abstract

Velocity structure of the lithosphere-asthenosphere system, to the depth of about 350 km, is obtained for almost 400 cells, sized 1°×1° in the Mediterranean region. The models are obtained by the following sequence of methods and tools: surface-wave dispersion measurements and collection; 2D tomography of dispersion relations; non-linear inversion of cellular dispersion relations; smoothing optimization method to select a preferred model for each cell. The 3D velocity model, that satisfies Occam razor principle, is obtained as a juxtaposition of selected cellular models. The reconstructed picture of the lithosphere-asthenosphere system evidences the, globally well known, asymmetry between the W- and E-directed subduction zones, attributed to the westward drift of the lithosphere relative to the mantle. Different relationship between slabs and mantle dynamics cause strong compositional differences in the upper mantle, as shown by large variations of seismic waves velocity, consistent with Polarized Plate Tectonics model.

## Introduction

The Mediterranean area is a complex zones from the geodynamic point of view. The area is located between European and African main plates with several intervening subduction zones, back-arc basins, areas with extended crust, mantle upraise, extensive past and present volcanism of different times, and small sub-plates with different stress regimes. Earth's structure has been explored on a variety of scales but seismic body- and/or surface-waves alone are not able to resolve the problem in details, nor at crustal and upper mantle depths. Only the inclusion of additional, independent information in the non-linear inversion of broadband regional surface-wave velocity maps may generate a comprehensive and relatively realistic picture of the Earth's lithosphere-asthenosphere system.

A 3D shear-wave velocity model to the depth of about 350 km is defined by means of the application of several methods and tools: collection of surface-wave dispersion measurements; 2D tomography on a grid sized 1°×1°; non-linear inversion of local dispersion relations; local smoothing optimization method; and juxtaposition of representative cellular models. The model is interpreted in the framework of the Polarized Plate Tectonics (Doglioni and Panza, 2015) that rather easily and naturally explains most of the identified geodynamic processes and tectonic units of the study area.

## Method

Most of the collected data are group-velocities measured by frequency-time analysis (Levshin et al., 1989) by Pontevivo and Panza (2001, 2006), Karagniani et al. (2002), Raykova et al. (2004), Raykova and Nikolova (2007) and El Gabry et al. (2013). Other published measurements were included to increase the density (lateral resolution) and penetration depth of the data. The dispersion curves along similar paths are averaged and the resulting data set for Rayleigh waves dispersions span over the period range from about 5 s to 80 s for group velocities and from 10 s to 150 s for phase velocities.

The 2D tomography based on the Backus–Gilbert method (Yanovskaya, 2001) was used to map the local values of the group and/or phase velocities that characterize horizontal (at a specific period) and vertical (at a specific grid knot) variations in the Earth's structure. The choice of the set of periods is based on the vertical resolution of the data, as determined by the partial derivatives of dispersion values with respect to the structural parameters (Panza, 1981). Check board or similar tests (Foulger et al., 2013) it is not necessary to be performed since the lateral resolution of the tomographic maps is defined as the average size of the equivalent smoothing area and its elongation.

Tomography results were used to compile local group- and phase-velocity dispersion curves for cells sized 1°×1°. Global tomography study of Ritzwoller and Levshin (1998) supplied long-period dispersion data so the period range of the constructed group-velocity curves was extended to 150 s, while local dispersion curves span over a varying period range, according to the availability of the data. Each local dispersion value is qualified by an error that is combination of measurement error and tomographic resolution.

The non-linear inversion method, called "hedgehog", was applied to process the cellular dispersion curves and to obtain the shear-wave velocity models in Mediterranean region for cells sized 1°×1°. The results obtained by Raykova and Panza (2006, 2011, 2015), Raykova and Nikolova (2007), Panza et al. (2007), Brandmayr et al. (2010) and ElGabry et al. (2013) were upgraded and refined, as well as results of the inversion for more than 60 new cells in western Mediterranean were added. The structure of each cell was modeled as a stack of 19 homogeneous elastic isotropic layers and the layers in the depth range from 3 – 10 km to about 350 km was replaced by a finite number of discrete parameters. Each cell has specific uppermost crustal layers and bottom layers of Poissonian material (common to all cells, according to Du et al., 1998), that are invariable during inversion. Resolving power of the tomographic data was improved using a priori independent information in the parameterization (Pontevivo and Panza, 2006). The applied inversion is a trial-and-error optimized Monte Carlo search and the dispersion curves were computed for each tested model. The model is accepted as solution, if the difference between the computed and measured values at each period is less than the relevant error and if the root mean square (r.m.s.) value for the whole dispersion curve is less than the given limit. The non-linear inversion is multi-valued and a set of models was obtained as the solution for each cell. The number of accepted structures is controlled by average r.m.s. value of the cellular dispersion curves and in general not exceeded 15 models per cell.

One representative model (with its uncertainties) for each cell is required in order to construct a 3D model of the studied area and to define the geotectonic meaning of the resulting structures. Local optimized smoothing, LSO, method (Boyadzhiev et al., 2008), which follows the Occam's razor principle, was used to define by a formalized criterion the cellular representative model. This method decreases the introduction of artificial vertical discontinuities in $V_S$ models between neighboring cells, minimizes the lateral velocity gradient, reduces the dependence of the final model from the predefined grid, and keeps the 3D structure as homogeneous as possible, even though well defined discontinuities can be introduced as independent a priori information. The models are represented as velocity vectors with equal size and the divergence between two models is defined as the standard Euclidean norm. The representative cellular model is the one, which minimizes the norm between models from neighboring cells. The direction of "maximum stability" is followed in the progressive choice of the cellular representative solution until the whole investigation domain is explored. Since the non-linear inversion and its smoothing optimization guarantee only the mathematical validity of the solution of the inverse problem, the optimization procedure may be repeated whenever necessary, including additional geophysical constrains.

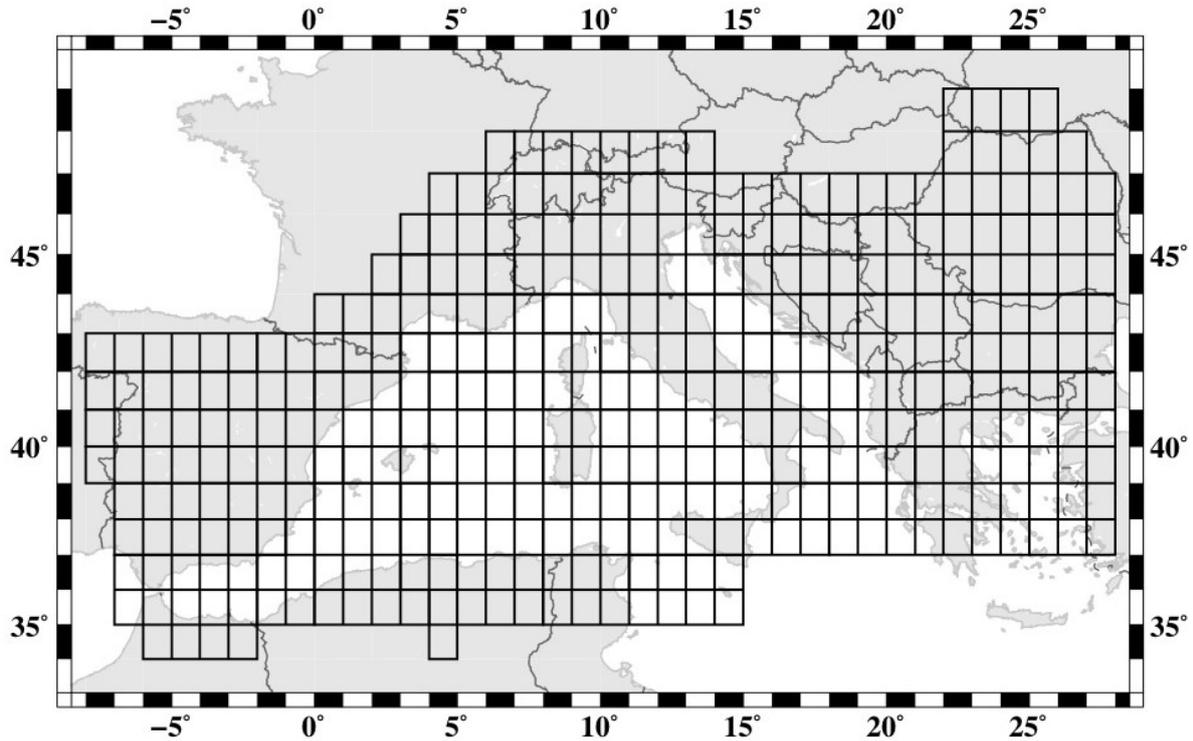

Fig. 1. Mediterranean region with studied cells represented by rectangles.

**Results**

About 400 cells have been processed in the Mediterranean region (Fig. 1). Independent information about the depth of Moho boundary (Grad et al., 2009; Tesauro et al., 2010) was used for the evaluation of the cellular models. Depth distribution of seismicity was used to resolve some ambiguities in layer's definition, using the revised ISC catalogue (ISC, 2007) for the period 1904–June 2005. Hypocenters were grouped in depth intervals (4 km for the crustal seismicity and 10 km for the mantle seismicity), consistent with the uncertainties of the hypocenter's depth calculation. LSO was run for three main parts: Balkans, Italy and Iberia, with some cells in common and keeping some of the cellular solutions fixed, according to previous studies (Brandmayr et al., 2010; Raykova and Panza, 2011 and 2015). The 3D shear-wave velocity structure of the study region is constructed as juxtaposition of the 1D representative cellular models. Each 1D structure provides the preferred layer's value of $V_S$ and thickness with uncertainties, in general equal to ±1/2 of the relevant parameter step used in the inversion.

A low velocity mantle layer is detected almost everywhere in the Mediterranean area with few exceptions under some of the orogenic belts, interpreted as lithospheric roots or subducted lithosphere. Transition from the lithosphere to the asthenosphere for most of the Mediterranean cells is sharp ($\Delta V_S > 0.20$ km/s) that could not be explained by classical temperature gradient (Rychert, 2015). The presence of water and melt that weaken the mantle is one of the hypotheses for such $V_S$ gradients. Part of these gradients should be associated also with anisotropy caused by directional dependence of seismic-wave velocity, produced by global mantle circulation. The Tyrrhenian Sea is the hottest zone in Mediterranean with numerous active volcanoes. It is quite similar to the asymmetric Lau Basin (Tonga ridge zone, mirrored by equator) studied by Shawn Wei et al. (2015). Similarly, the velocity models indicate that erupting sources are fed by an upwelling zone originating from the ambient mantle to the west of the back-arc basin (as detected by Panza et al., 2007 for Tyrrhenian Sea), well far from the subducting zones. Anomaly variations result from changes in the efficiency of melt extraction with decrease in melt, to the north in Tyrrhenian and to the south in the Lau basin, correlated with increased fractional melting and higher water content in the magma. Water released from the slabs may greatly reduce the melt viscosity or increase grain size, thereby facilitating melt transport.

Carpathians and Apennines have similar evolution, being Neogene W-directed subduction zones that initiated along the retrobelt of two preexisting orogens, associated to E- or NE-directed subduction (Alps and Dinarides). Differently from the Tyrrhenian (partly floored by oceanic crust), the Pannonian back-arc basin is an inland zone with relevant magmatism and thinned crust. The Ionian-Adriatic plate subducts both W-ward under the Apennines, and NE-ward beneath the Eurasian plate (Dinarides-Hellenides). In the meantime, the Adriatic plate overrides NW-ward (SE-directed subduction). The Aegean basin can be inferred as a continental rifting generated by the faster SW-ward advancement of the Greece lithosphere over the African plate with respect to the Cyprus slower subduction zone, hence providing widespread extension in Anatolia and the Aegean areas. The rifting is likely associated to melt and magma assisted seismicity to the depths below 40 km, shown by the cellular seismicity-depth distributions.

**Conclusions**

A structural model for the lithosphere-asthenosphere system of the Mediterranean region is here proposed as VS-depth spatial distribution. The multidisciplinary approach was used to better constrain the obtained model to independent geological, geophysical and petrological information. The reconstructed picture of the lithosphere-asthenosphere system evidences the globally observed asymmetry between the W- versus the E or NE-directed subduction zones, attributed to the westward drift of the lithosphere relative to the mantle.

Different relationship between slabs and mantle dynamics cause strong compositional differences in the upper mantle, as shown by large variations of seismic waves velocity. The complex features in the Mediterranean region can be easily framed within the Polarized Plate Tectonics model (Doglioni and Panza, 2015).

**References**

Boyadzhiev G., Brandmayr E., Pinat T., Panza G.F., 2008, Optimization for non-linear inverse problem, Rendiconti Lincei: Sci. Fis. e Nat., 19, 17-43.

Brandmayr E., Raykova R.B., Zuri M., Romanelli F., Doglioni C., Panza G.F., 2010, The lithosphere in Italy: structure and seismicity. In: M. Beltrando, A. Peccerillo, M. Mattei, S. Conticelli, and C. Doglioni (Eds.), J. Virtual Expl., 36 (paper 1), doi:10.3809/jvirtex.2009.00224.

Doglioni C. and Panza G.F., 2015, Polarized Plate Tectonics. Advances in Geophysics, Elsevier, 56, pp1-167, ISSN: 0065-2687.

Du Z.J., Michelini A., Panza G.F., 1998, EurID: a regionalized 3D seismological model of Europe, Phys. Earth Planet. Inter., 105, 31-62.

Grad M., Tiira T., 2008, The Moho depth map of the European Plate, Geophys. J. Int., 176, 279-292.

ElGabry M., Panza G.F., Badawy A., Korrat I., 2013, Imaging a relic of complex tectonics: the lithosphere-asthenosphere structure in the Eastern Mediterranean, Terra Nova, 25 (2), 102–109.

Foulger G.R., Panza G.F., Artemieva I.M., Bastow I.D., Cammarano F., Evans J.R., Hamilton W.R., Julina B.R., Lustrino M., Thybo H., Yanovskaya T.B., 2013. Caveats on tomographic images, Terra Nova, 25, 259-281.

ISC, 2007, International Seismological Centre, http://www.isc.ac.uk.

Karagianni E.,Panagiotopoulos D.G., Panza G.F., Suhadolc P., Papazachos C.B., Papazachos B.C., Kiratzi A., Hatzfeld D., Makropoulos K., Priestley K., Vuan A., 2002, Rayleigh Wave Group Velocity Tomography in the Aegean area, Tectonophysics, 358, 187-209.

Levshin A.L., Yanovskaya T.B., Lander A.V., Bukchin B.G., Barmin M.P., Its E.N., Ratnikova L.I., (ed. V.I. Keilis-Borok), 1989, Seismic Surface Waves in Laterally Inhomogeneous Earth, Kluwer Publ., Dordrecht.

Panza, G.F., 1981. The resolving power of seismic surface waves with respect to crust and upper mantle structural models. In: Cassinis R (ed) The solution of the inverse problem in geophysical interpretation. Plenum Press New York, 39-77.

Panza G.F., Raykova R.B., Carminati E., Doglioni C., 2007, Upper mantle flow in the western Mediterranean, Earth Planet. Sci. Lett., 257, 200–214.

Pontevivo A., Panza G.F., 2001, Group velocity tomography and regionalization in Italy and bordering areas, ICTP preprint IC/2001/136, Miramare, Trieste, 35 pp.

Pontevivo A., Panza G.F., 2006, The lithosphere-asthenosphere system in the Calabrian Arc and surrounding seas - Southern Italy, Pure Appl. Geophys., 163, 1617-1659.

Raykova R.B., Chimera G., Farina B., Panza G.F., 2004, S-wave velocity structure of the lithosphere-asthenosphere system in Mediterranean region, 32nd Int. Geol. Congr, Abs. Vol., pt. 2, abs. 208-5, p. 970.

Raykova R.B., Panza G.F., 2006, Surface wave tomography and non-linear inversion in the southeast Carpathians, Phys. Earth Planet. Interior., 157, 164–180.

Raykova R.B., Nikolova S.B., 2007, Tomography and velocity structure of the crust and uppermost mantle in southeastern Europe obtained from surface wave analysis, Stud. Geophys. Geod., 51, 166–184.

Raykova, R. B., Panza G. F., 2011, The shear-wave velocity structure of the lithosphere-asthenosphere system in the Iberian area and surroundings. Rendiconti Lincei - Scienze Fisiche e Naturali, 21,183-231.

Raykova, R. B., Panza G. F., 2015, VS structure of the crust and upper mantle in the Balkan Peninsula region. Book of Abstracts, 7th National Geophysical Conference, Sofia, Bulgaria, pp4.

Ritzwoller M., Levshin A.L., 1998, Eurasian surface wave tomography: group velocities, J. Geophys. Res., 103, 4839-4878.

Rychert, C., 2015, The slippery base of a tectonic plate, Nature, 518, 39-40.

Tesauro M., Kaban M.K., Cloetingh S.A.P.L., 2008, EuCRUST-07: a new reference model for the European crust, Geophys. Res. Lett., 35.

Shawn Wei S., Wiens D, Zha Y., Plank T, Blackman D., Dunn R., Conder J., 2015. Seismic evidence of effects of water on melt transport in the Lau back-arc mantle. Nature, 518, 395-398.

Yanovskaya T.B. (2001) Development of methods for surface-wave tomography based on Backus-Gilbert approach. In: Keilis-Borok V, Molchan GM (eds) Computational seismology. Am Geophys Union 32, Washington, DC, pp 11–26